\documentclass{jkas}

\def\year{2017} 
\def\month{June} 
\def\volume{50} 
\def\issue{3} 
\def\beginpage{1} 
\def\endpage{11} 
\def\received{May 9, 2017} 
\def\accepted{June **, 2017} 
\date{Received \received; accepted \accepted}

\def\kms{~{\rm km~s^{-1}}}
\def\cm3{~{\rm cm^{-3}}}
\def\yr{~{\rm yr}}
\def\Myr{~{\rm Myr}}
\def\muG{{\mu\rm G}}
\def\Mpc{~{\rm Mpc}}
\def\kpc{~{\rm kpc}}
\def\arcsec{^{\prime \prime}}

\usepackage{flushend}

\title{Shock Acceleration Model with Postshock Turbulence for Giant Radio Relics}

\author{Hyesung Kang}

\affil{Department of Earth Sciences, Pusan National University, Busan 46241, Korea; \email{hskang@pusan.ac.kr}}

\def\corrauthor{H. Kang}

\def\runningauthor{Kang}

\def\runningtitle{Shock Acceleration Model for Radio Relics}

\def\keywords{acceleration of particles --- cosmic rays ---
galaxies: clusters: general --- shock waves}

\def\abstracttext{
We explore the shock acceleration model for giant radio relics, in which relativistic electrons are 
accelerated via diffusive shock acceleration (DSA) by merger-driven shocks in the outskirts 
of galaxy clusters.
In addition to DSA, turbulent acceleration by compressive MHD mode 
downstream of the shock is included as well as energy losses of postshock electrons
by Coulomb scattering, synchrotron emission, and inverse Compton scattering off the cosmic background radiation.
Considering that only a small fraction of merging clusters host radio relics,
we favor the reacceleration scenario in which radio relics are generated preferentially 
when shocks encounter the regions containing low-energy ($\gamma_{\rm e} \lesssim 300$) cosmic ray electrons (CRe).  
We perform time-dependent DSA simulations of spherically expanding shocks with physical parameters relevant for
the Sausage radio relic, and calculate the radio synchrotron emission 
from the accelerated CRe.
We find that significant level of postshock turbulent acceleration is required in order to reproduce broad profiles 
of the observed radio flux densities of the Sausage relic. 
Moreover, the spectral curvature in the observed integrated radio spectrum can be explained, 
if the putative shock should have swept up and exited out of
the preshock region of fossil CRe about 10~Myr ago.
}

\begin{document}
\jkashead 

\section{Introduction}

The Sausage relic is a giant radio relic detected in the outskirts of the merging cluster CIZA
J2242.8+5301 located at the redshift, $z=0.188$ \citep{vanweeren10}. 
It is an arc-like radio structure whose spectral index increases
away from the edge of the relic toward the cluster center. Its volume-integrated radio spectrum has a power-law form 
with a steep spectral curvature above $\sim 2 $~GHz \citep{stroe13, stroe14}.
In \citet{vanweeren10}, the observed radio spectrum at the relic edge was interpreted as
a power-law with the slope, $\alpha_{\rm sh}\approx 0.6$, which can be translated into the 'radio Mach number', 
$M_{\rm rad}=[(3+2\alpha_{\rm sh})/(2\alpha_{\rm sh}-1)]^{1/2} \approx 4.6$, 
based on the diffusive shock acceleration (DSA) model.
On the other hand, the Mach number inferred from the Suzaku X-ray observations of \citet{akamatsu15}
indicates a much lower `X-ray Mach number', $M_{\rm X}=2.7_{-0.4}^{+0.7}$.
However, recent radio observations by \citet{stroe16} showed that the spectral index between 153 and 608~MHz may be
fitted by $\alpha_{153}^{608}\approx 0.7$ slightly downstream of the hypothesized shock location, 
if we ignore the flattest point with $\alpha_{153}^{608}\approx 0.54$ near the relic edge (see Figure 3 below).
This gives a much lower radio Mach number, $M_{\rm rad}\approx 3.3$, which is more comparable to $M_{\rm X}$.

The spectral steepening at high frequencies in the observed integrated spectrum, $J_{\nu}$, of the Sausage relic is 
not consistent with a single power-law energy spectrum of relativistic electrons
that are expected to be accelerated by a steady planar shock \citep{stroe14, stroe16}.
So \citet{kangryu16} suggested that such a spectral curvature could be explained,
if the relic is generated by the shock that sweeps through and moves out of a finite-size cloud with preexisting
cosmic ray electrons (CRe).
Lack of seed electrons outside of the cloud results in softening of the volume-integrated
electron spectrum beyond radiative cooling alone.
Moreover, it was pointed out that the ubiquitous presence of radio galaxies, AGN relics and radio phoenix implies that
the intralcluster medium (ICM) may contain {\it fossil} CRe left over from radio jets
\citep{ensslin99,slee01,clarke13,pinzke13, deGasperin15,kang16a}.

In addition to the integrated spectrum,
radio flux density, $S_{\nu}$, can be used to constrain the shock model parameters such as the shock speed and magnetic field strength.
The transverse length scale of the radio relic at high frequencies, for instance, is related with
the cooling length of the electrons with synchrotron peak frequency, $\nu_{\rm peak} \approx 0.3 (3eB/4\pi{m_e c}) \gamma_e^2$ (in $cgs$ units):
\begin{eqnarray}
& &\Delta l_{\rm cool} = t_{\rm rad}(\gamma_e)\cdot u_{2,3}\nonumber\\
& &\approx 100\ {\rm kpc} \cdot u_{2,3}\cdot Q\cdot
\left[{{\nu_{\rm obs}(1+z) \over {0.63{\rm GHz}}} }\right]^{-1/2},
\label{lwidth2}
\end{eqnarray}
where $t_{\rm rad}$ is the radiative energy loss time scale of CRe, $u_{2,3}=u_2/10^3 \kms$ is the downstream flow speed,
$\nu_{\rm obs}$ is the observation frequency and $z$ is the redshift of the host cluster \citep{kang16a}.
Here, the factor $Q$ is defined as
\begin{equation}
Q(B,z)\equiv \left[ { {(5~\muG)^2} \over {B^2+B_{\rm rad}(z)^2}}\right] \left({B \over 5 ~\muG}\right)^{1/2},
\label{qfactor}
\end{equation}
where $B_{\rm rad}=3.24~\muG(1+z)^2$ takes account for energy losses due to inverse Compton (iC) scattering 
off the cosmic background radiation and $B$ is expressed in units of $\muG$ \citep{kang11}.
Besides $\Delta l_{\rm cool}$, the radio flux density profiles projected in the sky plane are also affected by 
the geometrical shape of
the downstream volume and the viewing orientation (see Figure 1 of \citet{kang15}).

In \citet{kang16b} (Paper I), we attempted to reproduce the observed radio flux density profiles and the integrated spectrum
by the reacceleration model in which a shock of the sonic Mach number, $M_s\approx 3$, sweeps through 
a finite-sized cloud with a preexisting population of CRe, $f_{\rm pre} \propto p^{-s} \exp [-(p/p_{e,c})^2]$
with $s=4.1$.
A few shortcomings of this scenario are (1) the preexisting CRe is required to have
a flat energy spectrum with high cutoff energy ($\gamma_{e,c}=p_{e,c}/m_e c\approx 3-5\times 10^4$),
(2) the dimension of the pre-shock region with such CRe should be as large as $\sim 400$~kpc across the width of the relic
and $\sim 2$~Mpc along the length of the relic,
and (3) the adopted temperatures of the preshock and postshock region,
$kT_1=3.4$~keV and $kT_2=10.7-13.1$~keV are higher than the observed values,
$kT_{\rm 1,obs}= 2.7_{-0.4}^{+0.7}$~keV and $kT_{\rm 2,obs}= 8.5_{-0.6}^{+0.8}$~keV  \citep{akamatsu15}, respectively.
Hereafter,the subscripts 1 and 2 identify the upstream and downstream
states of a shock, respectively.

The first and second requirements may be considered somewhat unrealistic, because CRe with $\gamma_{e,c}> 3\times 10^4$
cool radiatively in short cooling times ($t_{\rm rad} <30\Myr$) in $\muG$-level magnetic fields.
So it would be challenging to maintain or replenish CRe with such a flat energy spectrum over such a large preshock volume 
by radio jets or lobes from AGNs.

On the other hand, the third requirement for high temperatures (i.e., large $u_2$) is necessary to reproduce the broad length scale
of spectral steepening, $u_2 t_{\rm rad}\sim 150$~kpc, in the postshock flow \citep{donnert16}. 
Alternatively, we could increase the postshock cooling length by introducing an additional acceleration process
such as turbulent acceleration behind the shock \citep[e.g.,][]{kang17}.
In merging clusters, turbulence can be injected into the ICM and cascade down to smaller scales,
which may further energize relativistic electrons via stochastic Fermi II acceleration, resulting in radio halos \citep{brunetti2014}.
Similarly, MHD/plasma turbulence can be generated at collisionless shocks, which could lead to turbulent acceleration of CRe
in the postshock region of radio relics.
As in \citet{kang17}, here we will consider the electron interactions with the compressive fast mode of MHD turbulence via the transit time damping (TTD) resonance,
which is thought be the dominant process in the high beta ($\beta_p=P_g/P_B \sim 50-100$) ICM plasma \citep{brunetti2007, brunetti2011}.

In this study, we explore the reacceleration model for the Sausage relic, which is different from the models considered in Paper I 
in the following aspects:
(1) Preexisting CR electrons consist of only low energy electrons with $\gamma_e \lesssim 300$ that have long cooling times 
($t_{\rm rad} \gtrsim 3.5$~Gyr), so they merely provide seed electrons to be injected to the Fermi I process.
(2) The radio spectral index of the relic at the shock location is determined by the shock Mach number, 
$M_{\rm s}\sim 3$, 
instead of the energy  spectrum of preexisting CRe (i.e. $s$ and $\gamma_{e,c}$).
(3) The shock-accelerated electrons are further accelerated by the Fermi II process due to postshock turbulence,
delaying the spectral aging of CRe behind the shock.

In the next section, the numerical simulations and the shock models are described.
The comparison of our results with observations is presented in Section 3,
followed by a brief summary in Section 4.

\section{Numerical Calculations}

The numerical setup for our DSA simulations was described in detail in Paper I and \citet{kang17}.
Some basic features are repeated here in order to make this paper self-contained.

\subsection{DSA Simulations for 1D Spherical Shocks}

We assume that the Sausage relic can be represented by a wedge-like patch of a spherical shell
shown in Figure 1 of \citet{kang15}, whose depth along the line-of-sight is specified by the extension angle $\psi$.
The spherical shell that contains radio-emitting electrons is assumed to be generated
by a spherically expanding shock and its downstream volume.

We follow the electron acceleration by DSA at the shock, and radiative cooling and turbulent acceleration
in the postshock region by solving the diffusion-convection equation in the one-dimensional (1D) spherically symmetric geometry:
\begin{eqnarray}
& &{\partial g_{\rm e}\over \partial t} + u {\partial g_{\rm e} \over \partial r} \nonumber\\
& &= {1\over{3r^2}} {{\partial (r^2 u) }\over \partial r} \left( {\partial g_{\rm e}\over
\partial y} -4g_{\rm e} \right) 
+ {1 \over r^2}{\partial \over \partial r} \left[r^2 \kappa(r,p){\partial g_{\rm e} \over \partial r} \right] \nonumber\\
& &+ p {\partial \over \partial y}\left[ {D_{pp} \over p^3} \left( {\partial g_{\rm e}\over \partial y} -4g_{\rm e} \right) \right] 
+ p {\partial \over {\partial y}} \left( {b\over p^2} g_{\rm e} \right),
\label{diffcon}
\end{eqnarray}
where $f_e(r,p,t)=g_e(r,p,t)p^{-4}$ is the pitch-angle-averaged phase space distribution function
for CRe, $u(r,t)$ is the flow velocity, $y=\ln(p/m_e c)$, $m_e$ is the electron mass, and $c$ is
the speed of light \citep{skill75}.
Here $r$ is the radial distance from the cluster center.

We adopt a Bohm-like spatial diffusion coefficient, $\kappa(p)= \kappa_N \cdot (p/m_ec)$ for relativistic electrons, 
where the normalization factor, $\kappa_N= k_{\rm B} \cdot m_ec^3/(3eB)= k_{\rm B} \cdot 1.7\times 10^{19} {\rm cm^2s^{-1}}/B_{\muG} $, 
with $B_{\muG}$ expressed in units of $\muG$.
The numerical factor, $k_{\rm B}$, depends on the strength of turbulent magnetic fields, $\delta B$
and becomes $k_{\rm B}=1$ for Bohm diffusion that represents the particle diffusion in completely random fluctuating fields.
The electron energy loss term, $b(p)= \dot p_{\rm Coul} + \dot p_{\rm sync+iC} $, accounts for Coulomb scattering, 
synchrotron emission, and iC scattering off the cosmic background radiation \citep[e.g.,][]{sarazin99}.

Here we explore a scenario in which the postshock electrons gain energy from turbulent waves via Fermi II acceleration,
thus abating spectral aging downstream of the shock. 
As in \citet{kang17}, we consider a simple model based on TTD resonance with compressive 
fast-mode MHD turbulence, since that is likely to be the most efficient turbulent acceleration process in the ICM \citep{brunetti2007, brunetti2011}.
The momentum diffusion coefficient for TTD resonance can be modeled as
\begin{equation}
D_{pp} = { p^2 \over {4\ \tau_{\rm acc}}}, 
\label{Dpp}
\end{equation}
where $\tau_{\rm acc}$ is an effective acceleration time scale for turbulent acceleration. 
In order to model the decay of turbulence behind the shock, 
we assume the turbulent acceleration time increases behind the shock on the scale of $r_{\rm dec}$ as
\begin{equation}
\tau_{\rm acc} = \tau_{\rm acc,0} \cdot \exp\left[{{(r_s-r)} \over r_{\rm dec}} \right]
\end{equation}
with, in most of our simulations, $\tau_{\rm acc,0}\approx 10^8 \yr$ and $r_{\rm dec}\approx 100$ kpc.

\begin{figure}[t!]
\centering
\includegraphics[trim=1mm 4mm 4mm 8mm, clip, width=84mm]{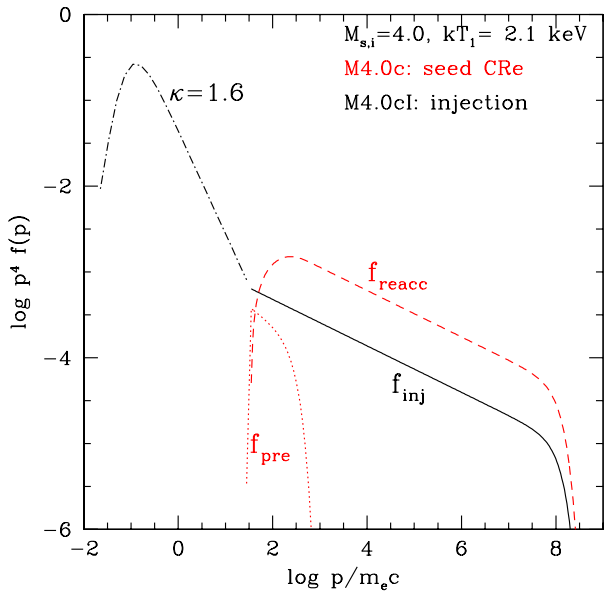}
\caption{Electron distribution function at the shock position, $f_{\rm inj}$ (black solid line) in the {\it in situ} injection model
and $f_{\rm reacc}$ 9red dashed) in the reacceleration model. 
The red dotted line shows the spectrum of preexisting seed CRe, $f_{\rm pre}$, while the black dot-dashed line shows the
$\kappa$-distribution with $\kappa=1.6$.
}
\end{figure}

\subsection{Injection versus reacceleration Model}

In this study, we consider the two kinds of DSA models: 
(1) in the {\it injection model}, suprathermal particles are generated via plasma kinetic 
processes near the shock and injected into the Fermi I process at the shock, and
(2) in the {\it reacceleration model}, preexisting low energy CRe are injected into the Fermi I process. 
In both models, the injected electrons are accelerated via DSA at the shock, and then they cool radiatively while being 
accelerated via turbulent acceleration behind the shock.

In the {\it injection model}, the electron population injected {\it in situ} from the background suprathermal population 
and then accelerated by DSA is modeled as
\begin{equation}
f_{\rm inj}(r_s, p)= f_N \left({p \over p_{\rm inj}}\right)^{-q} \exp \left[-\left({p \over p_{\rm eq}}\right)^2\right],
\label{finj}
\end{equation}
where $f_N$,  $q = 4M_{\rm s}^2/(M_{\rm s}^2-1)$, $p_{\rm inj}$, and $p_{\rm eq}$ are the normalization factor, 
the standard test-particle DSA power-law slope,
the injection momentum, and the cutoff momentum, respectively.
The injection momentum represents the low momentum boundary above which particles have mean free paths large enough to cross the
shock transition and thus participate in the Fermi I acceleration process.
Here, we adopt a simple model in which the electron injection depends on the shock strength as
$p_{\rm inj}\approx (6.4/\sigma) m_p u_s$ (where $\sigma=\rho_2/\rho_1$ is the shock compression ratio), in effect, resulting in $p_{\rm inj}\sim 150 p_{\rm th,e}$.
The cutoff momentum can be estimated from the condition that the DSA acceleration rate
balances the synchrotron/iC loss rate \citep{kang11}.
For typical parameters for the ICM shocks, $u_s\sim 3\times 10^3 \kms$ and $B_1\sim 1~\muG$, if we assume Bohm diffusion,
it becomes $p_{\rm eq}/m_{\rm e} c\sim 10^8$.

\begin{table*}
\begin{center}
{\bf Table 1.}~~Model Parameters for the Sausage Radio Relic\\
\vskip 0.3cm
\begin{tabular}{ lrrrrrrrrrlc }
\hline\hline

Model & $M_{\rm s,i}$ & $kT_1$ & $B_1$ & $L_{\rm cloud}$& $t_{\rm exit}$ & $t_{\rm obs}$& $M_{\rm s,obs}$ & $kT_{\rm 2,obs}$ & $u_{\rm s,obs}$& $N$& remarks\\
 {}  & &{(keV)} &($\muG$) & (kpc)& (Myr)& (Myr)& &{(keV)} & {(${\rm km~s^{-1}}$)}&$$  & \\

\hline

M3.5a  & 3.5 &2.5& 1& 420 & 144 & 155 & 2.97  & 9.0 & $2.4\times10^3$ & $1.6\times 10^{-4}$ & seed CRe\\
M3.5b  & 3.5 &2.5& 1& 485 & 167 & 177 & 2.93  & 8.8 & $2.4\times10^3$ & $1.6\times 10^{-4}$ & seed CRe\\
M3.5c  & 3.5 &2.5& 1& 581 & 200 & 214 & 2.86  & 8.5 & $2.3\times10^3$ & $1.6\times 10^{-4}$ & seed CRe \\
\hline
M4.0a   & 4.0 &2.1& 1& 451 & 144 & 159 & 3.34  & 9.1 & $2.5\times10^3$ & $1.2\times10^{-4}$ & seed CRe\\
M4.0b   & 4.0 &2.1& 1& 520 & 167 & 180 & 3.28  & 8.9 & $2.4\times10^3$ & $1.2\times10^{-4}$ & seed CRe\\
M4.0c   & 4.0 &2.1& 1& 624 & 200 & 211 & 3.21  & 8.6 & $2.4\times10^3$ & $1.2\times10^{-4}$ & seed CRe\\
\hline
M4.0cI   & 4.0 &2.1& 1& - & 200 & 211 & 3.21  & 8.6 & $2.4\times10^3$ & $\kappa=1.6$ & injection\\
M4.0cB   & 4.0 &2.1& 2.5& 624 & 200 & 211 & 3.21  & 8.6 & $2.4\times10^3$ & $1.2\times10^{-4}$ & stronger B\\
M4.0cN   & 4.0 &2.1& 1& 624 & 200 & 211 & 3.21  & 8.6 & $2.4\times10^3$ & $1.2\times10^{-4}$ & No TA\\

\hline
\end{tabular}
\end{center}
{$M_{\rm s,i}$: initial shock Mach number at the onset of the simulations ($t_{\rm age}=0$)}\\
{$kT_1$: preshock temperature}\\
{$B_1$: preshock magnetic field strength}\\
{$L_{\rm cloud}$: size of the cloud with preexisting CR electrons}\\
{$t_{\rm exit}$: shock age when the shock exit out of the cloud with preexisting electrons}\\
{$t_{\rm obs}$: shock age when the simulated results match the observations}\\
{$M_{\rm s,obs}$: shock Mach number at $t_{\rm obs}$}\\
{$kT_{\rm 2,obs}$: postshock temperature at $t_{\rm obs}$}\\
{$u_{\rm s,obs}$: shock speed at $t_{\rm obs}$}\\
{$N=P_{\rm CRe}/P_{\rm g}$: the ratio of seed CR electron pressure to gas pressure}\\
The subscripts 1 and 2 indicate the preshock and posthoock quantities, respectively.

\end{table*}

The factor $f_N$ depends on the suprathermal electron population in the background plasma, which is assumed to
be energized via kinetic plasma processes at the shock and
form a $\kappa$-distribution, rather than a Maxwellian distribution \citep{pierrard10}.
The value of $\kappa$ index is expected to depend on the shock parameters such as the obliquity angle
and the sonic and Alv\'evnic Mach numbers, in addition to the plasma parameters of the background medium.
For instance, the electron energy spectrum measured in the interplanetary medium near the Earth orbit can be 
fitted with the $\kappa$-distribution with $\kappa\sim 2-5$ \citep{pierrard10}. 
Here we adopt a somewhat flatter value of $\kappa\sim 1.6$ to maximize the electron injection rate.  
Figure 1 illustrates the $\kappa$-distribution (dot-dashed line) for $p< p_{\rm inj}$ and $f_{\rm inj}(r_s, p)$
(solid line) for  $p> p_{\rm inj}$ in one of the models considered below.
For $p\gg p_{\rm th,e}$, the $\kappa$-distribution can be approximated as 
$f_{\rm \kappa} \propto p^{-2(\kappa+1)}$, so the amplitude $f_N$ becomes smaller for a larger value of $\kappa$.
For example, the factor $f_N$ for $\kappa=2.5$ is smaller by a factor of about 200 than that for $\kappa=1.6$.

In the case of the {\it reacceleration model}, the preexisting seed CRs are assumed to have a power-law spectrum
with exponential cutoff as follows:
\begin{equation}
f_{\rm pre}(p) = f_o \cdot p^{-s} \exp \left[ - \left({p \over p_{e,c}} \right)^2 \right],
\label{fpre}
\end{equation}
where the slope $s=4.6$ and the cutoff $\gamma_{e,c}=300$ are adopted for all models considered here (see the red dotted line in Figure 1).
But the exact shape of $f_{\rm pre}(p)$ is not important, because the only significant role of these  
low-energy CRe is to provide seed particles to be injected to the DSA process.
Note that the electrons with $\gamma_{e,c}<300$ cool on the time scales longer than 3.5~Gyr, 
so they could represent fossil electrons in the ICM that are left over from AGN jets ejected early on.
The normalization factor, $f_o$, can be parameterized with the ratio of the preexisting CRe pressure 
to the gas pressure in the preshock region, 
$N \equiv P_{\rm CRe,1}/P_{\rm gas,1} \propto f_o$ for a given set of $s$ and $p_{e,c}$.
In the models considered here, typically, the models with $N \sim 10^{-4}$ produce the radio flux profiles that
can match the amplitude of observed flux in the Sausage relic.

The reaccelerated population of $f_{\rm pre}(p)$ at the shock can be calculated by
\begin{equation}
f_{\rm reacc}(r_s, p)= q \cdot p^{-q} \int_{p_{\rm inj}}^p p^{\prime q-1} f_{\rm pre} (p^\prime) dp^\prime 
\label{freacc}
\end{equation}
\citep{dru83}.
Note that if the DSA slope, $q$, is flatter (smaller) than the slope, $s$, of the preexisting population,
the downstream energy spectrum does not have any memory of the upstream spectrum other than its amplitude.
As can be seen in the red dashed and black solid lines in Figure 1, 
both $f_{\rm reacc}(r_s, p)$ and $f_{\rm inj}(r_s, p)$ have
the same power-law form.

Since the time scale for DSA at the shock is much shorter than the electron cooling time scale,
we can assume that electrons are accelerated almost {\it instantaneously} to $p_{\rm eq}$ at the shock front.
Moreover, the minimum diffusion length scale to obtain converged
solutions in simulations for diffusion-convection equation is much smaller than the typical downstream cooling length of $\sim 100$ kpc.
Taking advantage of such disparate scales, we adopt {\it analytic} solutions for the electron spectrum {\it at the shock location} as
$f(r_s,p)= f_{\rm inj}(r_s, p)$ or $f_{\rm reacc}(r_s, p)$,
while Equation (\ref{diffcon}) is solved outside the shock.
So, basically we follow the energy losses and turbulent acceleration of electrons behind the shock,
while the DSA analytic solutions are applied to the zone containing the shock. 

\subsection{Shock Parameters}

\begin{figure*}[t!]
\vskip -0.5cm
\centering
\includegraphics[trim=2mm 2mm 2mm 2mm, clip, width=150mm]{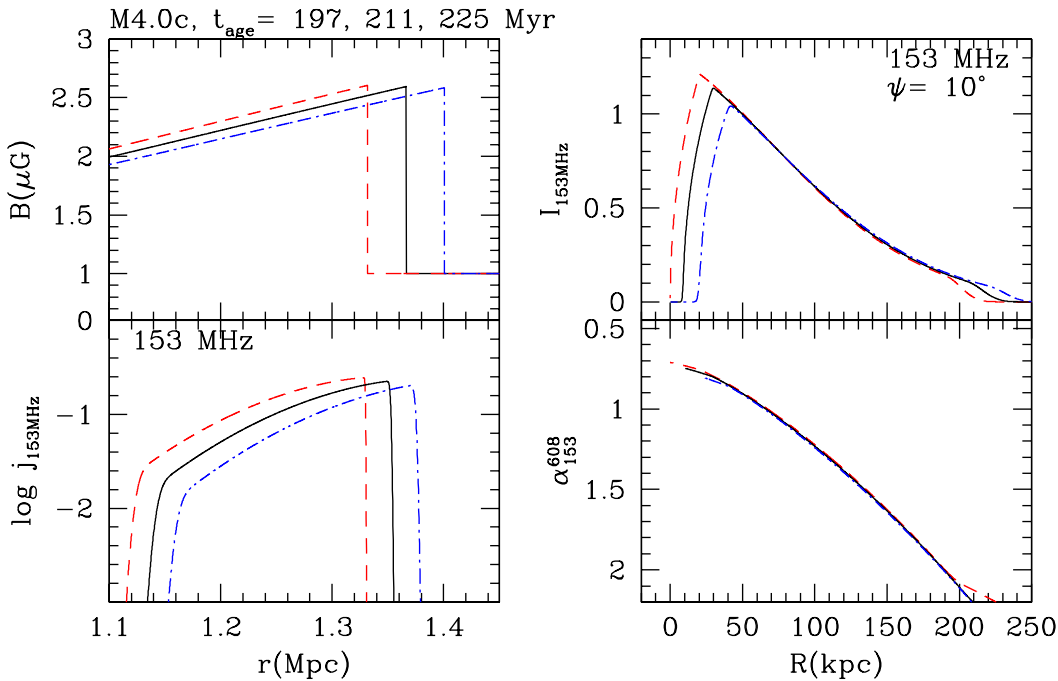}
\vskip -5.0cm
\caption{Results for the fiducial model M4.0c at $t_{\rm age}=$~197 (red dashed lines), 211 (black solid), 
and 225~Myr (blue dot-dashed). 
Left: magnetic field strength, $B(r)$ and synchrotron emissivity, $j_{\rm 153MHz}(r)$, where $r$ is the radial distance from the cluster center in units of Mpc.
Right: surface brightness, $I_{\rm 153MHz}(R)$, and spectral index, $\alpha_{153}^{608}$ between 153 and 608~MHz, where
$R$ is the projected distance behind the shock in units of kpc.
}
\end{figure*}

It is not well understood how merger-driven shocks evolve dynamically as they propagate in the cluster periphery.
In a major binary merger, shocks are launched after core passage of the two subclumps 
and propagate beyond the virial radius of the newly formed cluster \citep{vanweeren11}.
It is expected that in general shock speeds increase during the initial launch period and
may decrease later during the expansion stage.
In a realistic cluster merger, however, the merger is likely to involve subsequent infall of more subclumps along the
filaments connected with the cluster. So the dynamical evolution of a merger shock can be quite complex
\citep[e.g.][]{paul11}.

Here we assume that the shock dynamics can be approximated by a self-similar blast wave
that propagates through the isothermal ICM with the density profile of $n_{\rm H} = 10^{-4}\cm3 (r/0.8{\rm Mpc})^{-2}$,
where $n_{\rm H}$ is the number density of hydrogen atom.
So the shock radius and velocity evolves roughly as $r_s\propto t^{2/3}$ and $u_s\propto t^{-1/3}$,
respectively, where $t$ is the time since the point explosion for the spherical blast wave \citep[e.g.,][]{ryu91}.
During the simulation time period of $\sim 200$~Myr, the model shock speed decreases by a factor of $\lesssim 1.3$.

The ICM temperature upstream and downstream of the relic edge is observed to be
$kT_1 = 2.7_{-0.4}^{+0.7}$~keV and $kT_2 = 8.5_{-0.6}^{+0.8}$~keV, respectively,
which indicates the sonic Mach number of $M_s\approx 2.7$ \citep{akamatsu15}.
Since the shock speed decreases in time in our model, we consider two values for the initial
shock Mach number, $M_{\rm s,i}= 3.5$ and $4.0$, and two values for the preshock temperature,
$kT_1= 2.5$~keV and 2.1~keV, respectively.
Table 1 shows the model parameters for the DSA simulations considered here.

According to \citet{akamatsu15}, the discontinuity in the X-ray temperature distribution agrees well with the outer edge of the Sausage relic within the angular resolution of the {\it Suzaku} X-ray observation (2 arcmin$\approx 384$~kpc). In the case of the Toothbrush relic, on the other hand, the spatial offset of $\sim 1$ arcmin between the X-ray shock and the relic edge was indicated in the XMM-Newton observation by \citet{ogrean13}.
However, such discrepancy was rebutted by \citet{vanweeren16} where the refitted XMM Newton and the {\it Chandra} profiles were shown to be consistent with the radio flux profile of relic B1 of the Toothbrush relic.

At the onset of the simulations ($t=t_i$), the initial shock speed, $u_{\rm s,i}$ is specified by $M_{\rm s,i}$ 
and $kT_1$, while the shock location is assumed to be $r_{\rm s,i}=0.8$~Mpc.
This fixes the initial time $t_i$ when the shock encounters the cloud of preexisting CRe, 
and the scaling factors for the similarity solution, $\rho_o$, $u_o$, and $t_o$.
We define the ``shock age'', $t_{\rm age} \equiv t - t_i$, as the time since the onset of the simulations.

As in Paper I, in order to reproduce the spectral steepning about 2~GHz, we assume that, in the reacceleration model, at the onset of the simulations the shock encounters a cloud of 
size $L_{\rm cloud}$ containing preexisting seed CRs, and then exits out of it at $t_{\rm exit}$. 
So the size $L_{\rm cloud}$ affects the postshock profiles of radio flux densities.
Then the `time of observation', $t_{\rm obs}$, is chosen when both the simulated brightness profiles and the integrated spectra
become consistent with the observations reported by \citet{stroe16}.
Between the exit time, $t_{\rm exit}$, and $t_{\rm obs}$, the shock sweeps the region devoid of 
preexisting CRe, which results in steepening of the volume-integrated electron energy spectrum.
As a result, the elapsed period of $(t_{\rm obs}-t_{\rm exit})\approx 10-15 \Myr$ controls the spectral curvature of the integrated radio spectrum.

The fiducial value of the preshock magnetic field strength is set to be $B_1= 1\muG$,
which is assumed to be uniform in the upstream region.
As in Paper I, the postshock magnetic field strength is modeled as $B_2(t)=B_1 \sqrt{1/3+2\sigma(t)^2/3}\approx 2.5-2.7\muG$,
which decreases slightly as the shock compression ratio, $\sigma(t)$, decreases in time in response to shock evolution.
For the downstream region ($r<r_s$), we assume a simple model in which the magnetic field strength scales with the 
gas pressure as $B_{\rm dn}(r,t)= B_2(t) \cdot [P(r,t)/P_{2}(t)]^{1/2}$,
where $P_{2}(t)$ is the gas pressure immediately behind the shock (see Figure 2).

We adopt the model naming convention in Table 1, where the number after the first letter 'M'
corresponds to $M_{\rm s,i}$.
This is followed by a sequence label (a, b, c) as the size of the cloud containing preexisting CRe, $L_{\rm cloud}$,
increases.
The M4.0c model is the reacceleration model with fossil CRe with $M_{\rm s,i}=4.0$ and $L_{\rm cloud}=624$~kpc.
The M4.0cI model is the injection model in which only {\it in situ} injection from background suprathermal electrons
is included, 
while the M4.0cB model adopts a preshock magnetic field strength higher than that of the rest of the models.
In M4.0cI, the {\it in situ} injection is turned on at the onset of the simulation, and then it is
turned off at $t_{\rm exit}=200 \Myr$ to create a spectral curvature at high frequencies.
In the M4.0cN model, turbulent acceleration is turned off to demonstrate its effects on the postshock spectral aging.

\begin{figure*}[t!]
\vskip -0.5cm
\centering
\includegraphics[trim=2mm 2mm 2mm 2mm, clip, width=160mm]{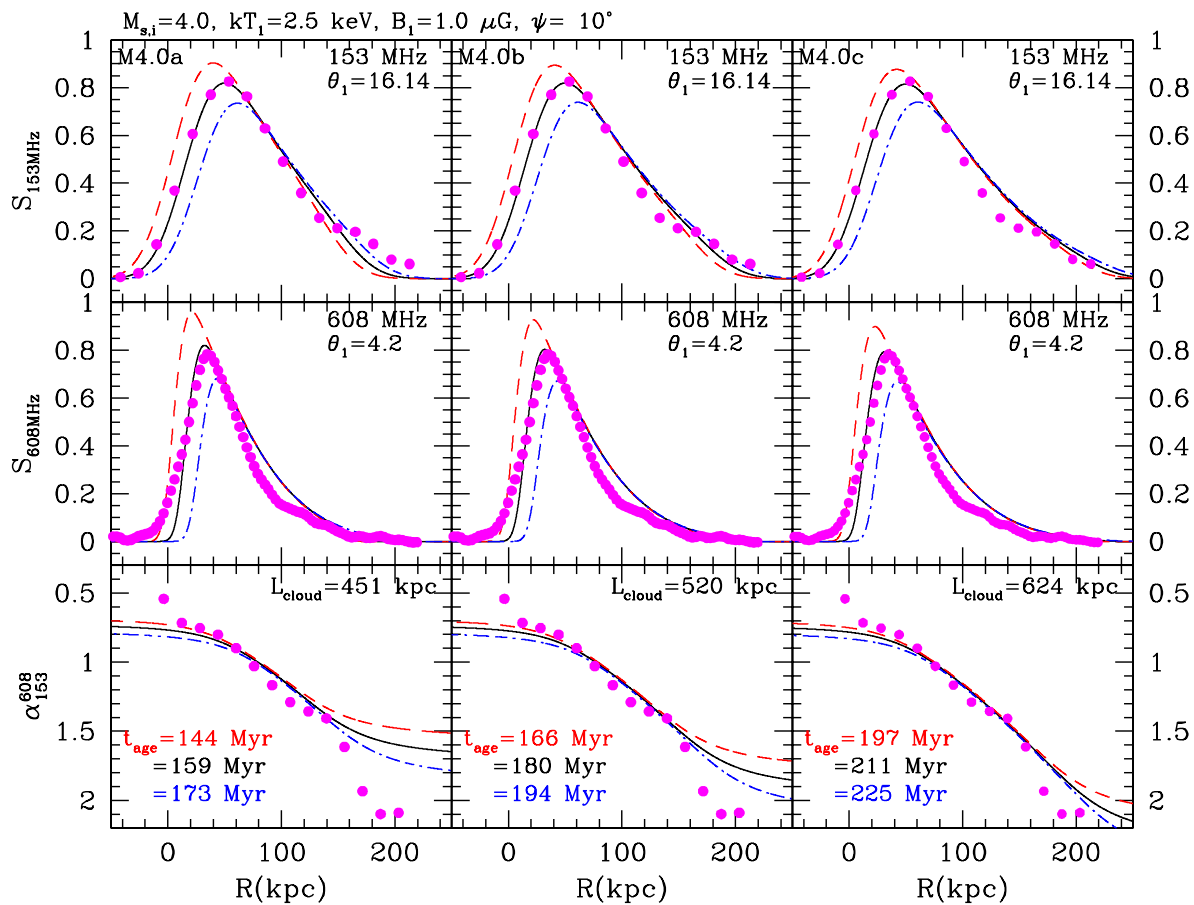}
\vskip -3.0cm
\caption{Beam convolved brightness profiles $S_{\nu}(R)$ at 153~MHz (top panels) and at 608~MHz (middle panels),
and the spectral index $\alpha_{153}^{608}$ between the two frequencies (bottom panels)
are plotted for three shock ages (red, black, and blue lines), specified in the bottom panels.
Here $R$ is the projected distance behind the shock in units of kpc.
The extension angle, $\psi=10^{\circ}$, is adopted.
The results are shown for the M4.0a model with $L_{\rm cloud}=451$~kpc (left-hand panels), 
M4.0b model with $L_{\rm cloud}=520\kpc$ (middle panels), and
M4.0c model with $L_{\rm cloud}=624\kpc$ (right-hand panels).
The simulated brightness profiles, $I_{\nu}(R)$, are smoothed with
Gaussian smoothing with $51.7\kpc$ (equivalent to the beam angle $\theta_1=16.14^{\arcsec}$) for 153~MHz
and with $13.4\kpc$ (equivalent to $\theta_1=3.42^{\arcsec}$) for 608~MHz to be compared with the observed 
flux profiles of \citet{stroe16} (magenta filled circles).
}
\end{figure*}

The eighth and ninth columns of Table 1 show the shock Mach number and the postshock temperature at $t_{\rm obs}$:
$M_{\rm s,obs}=2.9-3.3$ and $kT_{\rm 2,obs}=8.5-9.1$~keV, which are reasonably consistent with the X-ray observations
reported by \citet{akamatsu15}.

We note here that $M_{\rm X}$ inferred from X-ray observations could be lower than $M_{\rm radio}$ estimated
from radio spectral index,
since a radio relic may be associated with multiple shocks.
According to mock observations of cluster shocks formed in structure formation simulations,
X-ray observations tend to pick up shocks with lower $M_{\rm s}$ along a given line-of-sight,
while radio emissions come preferentially from shocks with higher $M_{\rm s}$ \citep[e.g.,][]{hong15}.
In the case of the Toothbrush relic, it was shown that $M_{\rm s}\approx3.0$ is required to reproduce the radio data, while 
the X-ray data indicate $M_{\rm X}\approx 1.2-1.5$ \citep{kang17}.

Finally, the eleventh column shows, $N\approx 10^{-4}$, the ratio of the preexisting CRe pressure to the upstream gas 
pressure that can generates radio flux densities consistent with the observations reported by \citet{stroe16}.

\section{Results of DSA Simulations}

Figure 2 shows the DSA simulation results for the M4.0c model at three epochs: just before the shock exits out of the cloud at
$t_{\rm exit}\approx 200$~Myr (red dashed lines), at the time of observation, $t_{\rm obs}=211$~Myr (black solid), 
and at $t_{\rm age}=225$~Myr (blue dot-dashed).
The upper left panel shows the profiles of the magnetic field strength, which contain a discontinuous jump
at the shock location, $r_{\rm s}(t)=1.33-1.4$~Mpc.
Note that the shock expands radially outward in the left-hand panels of Figure 2.

\subsection{Surface Brightness and Spectral Index Profiles}

\begin{figure*}[t!]
\vskip -0.8cm
\centering
\includegraphics[trim=2mm 2mm 2mm 2mm, clip, width=160mm]{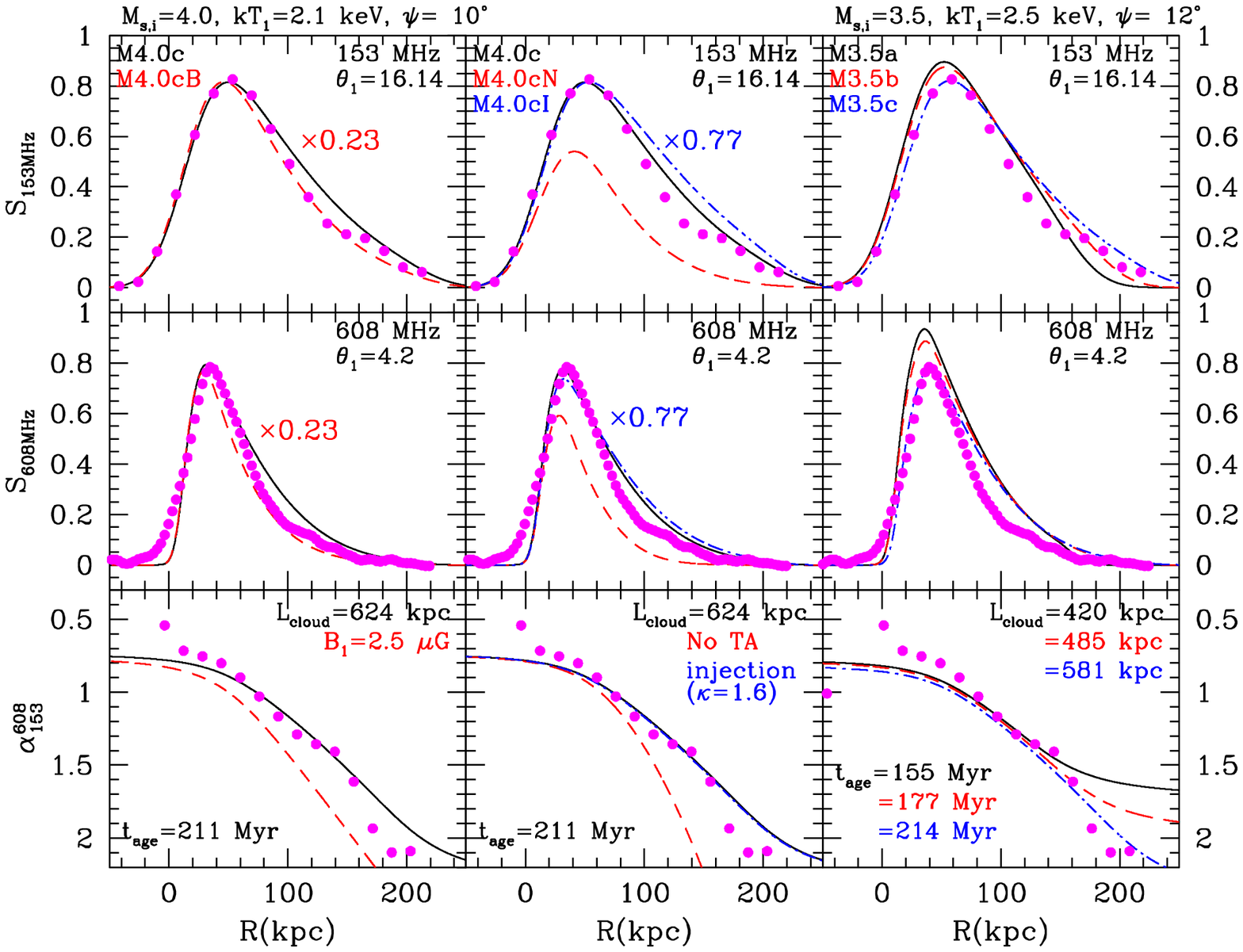}
\vskip -2.8cm
\caption{Same as Figure 2 except that the M4.0c and M4.0cB models with the extension angle $\psi=10^{\circ}$ are compared in the left-hand panels, 
the M4.0c, M4.0cN, and M4.0cI models with $\psi=10^{\circ}$ are compared in the middle panels,
and the M3.5a, M3.5b, and M3.5c models with $\psi=12^{\circ}$ are compared in the right-hand panels.
The radio flux density, $S_{\nu}$, is multiplied by a factor of 0.23 for M4.0cB and 0.77 for M4.0cI with respect to 
$S_{\nu}$ for the fiducial model, M4.0c.
}
\end{figure*}

\begin{figure*}[t!]

\centering

\includegraphics[trim=2mm 3mm 2mm 2mm, clip, width=160mm]{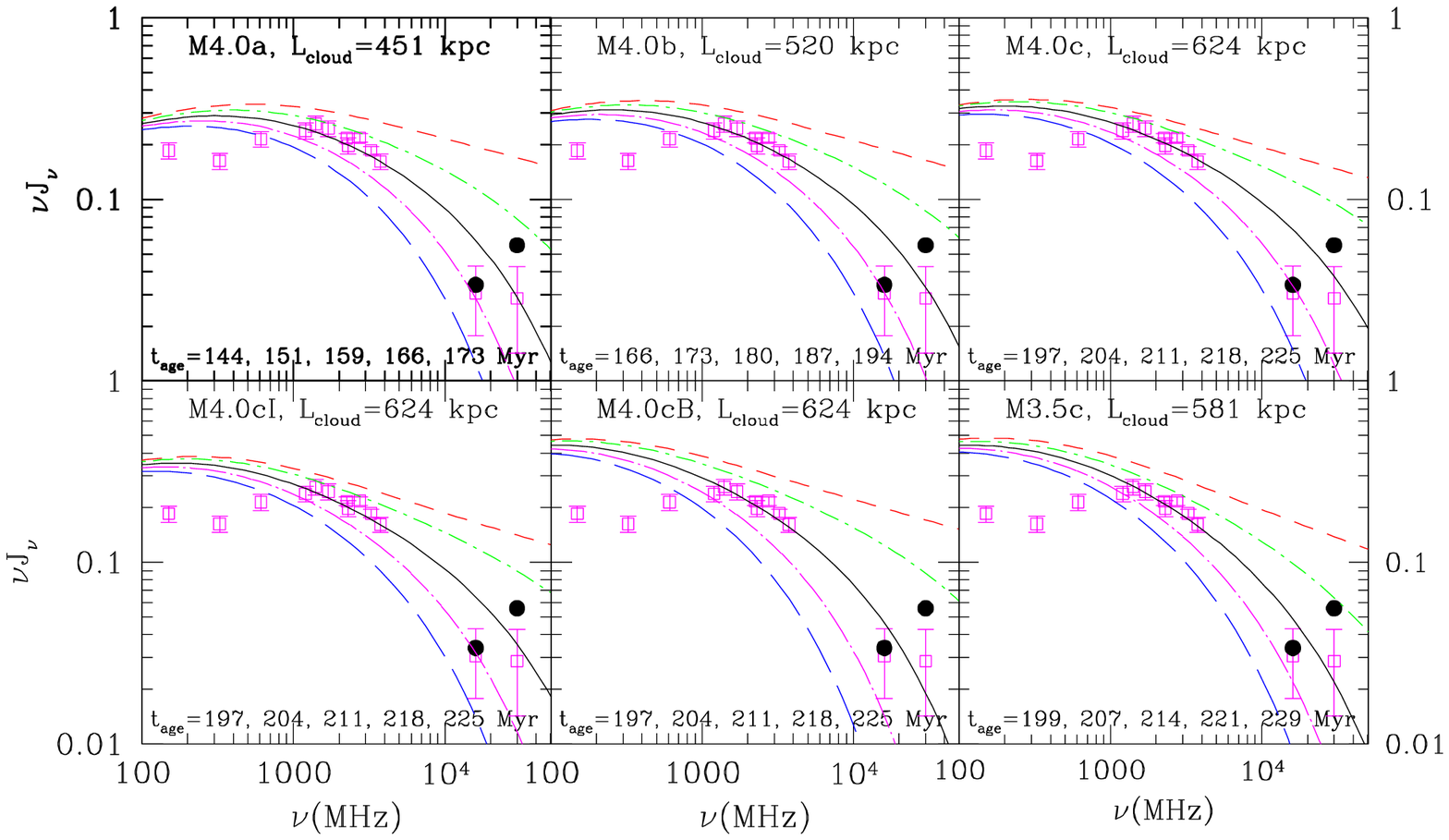}
\vskip -5.8cm
\caption{Time evolution of volume-integrated radio spectrum at five shock ages, specified in each panel,
are shown in chronological order by the red dashed, green dot-dashed, black solid, magenta dot-long dashed, and blue long dashed lines.
The open magenta squares and the error bars are the observational data taken from \citet{stroe16}.
The solid black circles} at 16~GHz and 30~GHz are the data points, multiplied by factors of 1.11 and 1.96, respectively, 
which could represent the SZ-corrected fluxes \citep{basu16}.
\end{figure*}
Using the CRe energy spectrum and the magnetic field strength in the model DSA simulations,
we first calculate the synchrotron emissivity $j_{\nu}(r)$ of each spherical shell. 
The lower left panel of Figure 2 demonstrates that the outermost edge of the synchrotron emissivity at 153~MHz, 
$j_{\rm 153MHz}(r)$, lags behind the shock location after the shock moves out of the cloud at $t_{\rm exit}$.

The radio surface brightness, $I_{\nu}(R)$, is calculated by integrating $j_{\nu}(r)$ along a given line-of-sight,
where a wedge-like postshock volume of radio-emitting electrons is adopted, as in Figure 1 of \citet{kang15}.
Here $R$ is the distance behind the projected shock edge in the plane of the sky.
This volume is specified by the extension angle, $\psi$, which is assumed to be about $10^{\circ}$ 
in the case of the Sausage relic \citep[e.g.,][]{vanweeren10,kang12}.
The upper right panel of Figure 2 shows $I_{\rm 153MHz}(R)$ at 153~MHz (in arbitrary units), using $\psi=10^{\circ}$.
Note that the shock faces to the left in the right-hand panels of Figure 2, so the region of $R<0$ is the preshock region.
Again, one can see that the edge of the radio relic is located the behind the shock (at $R=0$) after $t_{\rm exit}$.
The spectral index profile, $\alpha_{153}^{608}(R)$ in the lower right panel is calculated with the ratio between
$I_{\rm 153MHz}(R)$ and $I_{\rm 608MHz}(R)$.

In order to obtain beam-convolved flux density, the intensity $I_{\nu}(R)$ is smoothed by Gaussian smoothing with 51.7~kpc  
width (equivalent to $16.14^{\prime\prime}$) for 153~MHz and 13.4~kpc width (equivalent to $4.2^{\prime\prime}$) for 608~MHz.
For the profile of the spectral index, $\alpha_{153}^{608}(R)$, both $I_{\rm 153MHz}(R)$ and $I_{\rm 608MHz}(R)$
are smoothed with the same width of 51.7~kpc.

Figure 3 shows the time evolution of $S_{\rm 153MHz}(R)$, $S_{\rm 608MHz}(R)$, and $\alpha_{153}^{608}(R)$ in the M4.0a, b, c models
with different cloud size $L_{\rm cloud}$.
The times of observation, $t_{\rm obs}= 159,$ 180, and 211~Myr (black solid lines) are chosen for the M4.0a, b, c models, respectively.
At the times earlier (red dashed lines) or later (blue dot-dashed lines) than $t_{\rm obs}$ are shown for comparison.
Since the shock slows down and moves out of the cloud of preexisting CRe at $t_{\rm exit}$ (given in Table 1), 
the amplitude of $S_{\nu}$ decreases in time.

The observed flux density at 153~MHz for the beam of $16.14^{\prime\prime}\times 13.75^{\prime\prime}$ 
is $S_{\rm 153MHz}\approx 0.014$~Jy at $R\approx 55$~kpc \citep{stroe16}.
In Figure 3, the normalization factor of $S_{\nu}$ and its peak location are chosen so that the black solid 
lines match the observed data represented by the magenta solid circles.
The amount of preexisting CRe that matches the observed flux density corresponds to $N\approx 1.6\times 10^{-4}$ (see Table 1).

Although the three models can reproduce reasonably well both $S_{\rm 153MHz}(R)$ and $S_{\rm 608MHz}(R)$,
the M4.0c model (at 211~Myr) can fit best the observed profile of $\alpha_{153}^{608}(R)$.
So we take the M4.0c model as the `fiducial' model in this discussion.
In the M4.0a and M4.0b models, $L_{\rm cloud}$ is smaller, so the shock exits out of the cloud earlier, 
resulting in less spectral aging for $R>150$~kpc at $t_{\rm obs}$, compared to the M4.0c model
(see the bottom panels of Figure 3).
Of course, if we were to choose a later epoch for $t_{\rm obs}$ for these two models,
the spectral index profile would become more comparable to the observations.
As we will show in Figure 5 below, however, 
the time interval of $t_{\rm obs}-t_{\rm exit}$ becomes longer in those cases,
resulting in the spectral curvature of the integrated radio spectrum much steeper than observed.

Note that we do not attempt to fit the flattest data for $\alpha_{153}^{608}\approx 0.54$ at $R\approx0$~kpc.
This allows us to choose a much smaller shock Mach number, i.e., $M_{\rm s,obs}\approx 3.2$ for the M4.0c model,
instead of $M_{\rm radio}\approx 4.6$ suggested in earlier papers \citep[e.g.,][]{vanweeren10}.

The right-hand panels of Figure 4 show the results at $t_{\rm obs}=155,$ 177, and 214~Myr for M3.5a, b, c models, respectively.
The M3.5c model at $t_{\rm obs}=214$~Myr seems to give the best fit to the observations.
But the profiles of $\alpha_{153}^{608}(R)\gtrsim 0.8 $ for $R<80$~kpc are slightly steeper than the observed profile 
in these three models with $M_{\rm s,obs}=2.9-3.0$.

The left-hand panels of Figure 4 compare the M4.0cB (stronger $B_1$, red dashed lines) with the M4.0c model.
Stronger magnetic fields enhance the synchrotron emission and cooling, 
resulting in higher radio flux densities and a steeper profile of $\alpha_{153}^{608}(R)$.
So we reduce $S_{\nu}$ by a factor of 0.23 for the M4.0cB model in order to plot both models with the same normalization scaling.

In the middle panels of Figure 4, the M4.0cI (injection only, blue dot-dashed lines) and M4.0cN (no turbulent acceleration,
red dashed lines) are compared with the fiducial M4.0c model.
As mentioned in Section 2.2, the normalization for the injection model depend on the value of the $\kappa$ index 
for the suprathermal electrons in the background plasma (see Figure 1). 
With the adopted value $\kappa=1.6$ in M4.0cI, the peak value becomes $S_{\rm 153MHz}\approx 0.018~{\rm Jy}$,
so the radio flux densities are scaled down by a factor of $0.77$ for this model 
in order to compare them with the observational data in the Figure.
We note, however, a more realistic value would be $\kappa>2$, so the amplitude of $S_{\rm 153MHz}$ in the injection model
might be much smaller than observed.
In the M4.0cN model without turbulent acceleration,
$S_{\nu}$ is smaller and $\alpha_{153}^{608}$ is steeper, compared to the M4.0c model.

Figures 3 and 4 show that the predictions of the M4.0c model convolved with appropriate beam widths are in reasonable
agreement with the observations, providing that there exist fossil CRe with $N\approx 10^{-4}$ in the ICM.
This exercise demonstrates that
the profiles of observed radio flux density, $S_{\nu}(R)$, at multi frequencies can provide strong constraints 
on the model parameters for radio relics.

\subsection{Integrated Spectrum}

As shown in Paper I, the spectral curvature in the observed integrated radio spectrum of the Sausage relic
cannot be reproduced by a simple DSA model for a steady planar shock.
But it can be explained if we adopt an addition condition for a finite size of the cloud with 
preexisting CRe.
We note that in the {\it in situ} injection model (M4.0cI), the same kind of curvature can be created somewhat 
artificially by turning off the injection after $t_{\rm exit}=200 \Myr$.

Figure 5 shows the time evolution of the integrated spectrum, $\nu J_{\nu}$, for six different models.
The red dashed lines for each model show the spectrum at the first epoch just before the 
shock exits out of the cloud. 
They follow roughly the predictions based on the postshock radiative cooling, i.e.,
steepening of $J_{\nu}$ from $\nu^{-\alpha_{\rm s}}$ to $\nu^{-(\alpha_{\rm s}+0.5)}$ at $\sim$GHz.
Such description is only approximate here, because additional turbulent acceleration operates in the postshock region.

Then the green dot-dashed lines, black solid lines, magenta dot-long dashed lines, and blue long dashed lines 
present the spectra with progressively steeper curvatures at four later epochs in chronological order.
The open magenta squares and the error bars are observational data taken from Table 3 of \citet{stroe16}.
\citet{basu16} calculated that the amount of the Sunyaev-Zel’dovich (SZ) decrement in the observed radio flux 
for several well-known radio relics, based on models for the ICM electron density profile and the radio flux profile.
Although we know such predictions depend sensitively on those model details,
we adopt their estimates for the SZ contamination factor for the Sausage relic given in their Table 1.
Then, the SZ correction factors, $F$, for the fluxes at 16 GHz and 30 GHz are about 1.1 and 1.96, respectively. 
The two solid circles in each panel of Figure 5 correspond to the flux levels so-corrected at the two highest frequencies.

Note that \citet{stroe16} suggested that the observed spectrum could be fitted by a broken power-law: 
$\alpha=0.90\pm 0.04$ below 2~GHz and
$\alpha=1.77\pm 0.13$ above 2~GHz.
The black solid lines at $t_{\rm obs}$ for each model are chosen as the best fits to the observed spectrum 
in the range of $1-3$~GHz.
All six models seem to generate similar integrated spectra, 
although the simulated profiles of $\alpha_{153}^{608}(R)$ are rather different as shown in Figures 3 and 4.
Considering that the observation errors in the flux data is only 10 \% for $\nu \lesssim 3$~GHz,
it seems somewhat difficult to fit very well the observational data both below and above 1~GHz simultaneously 
with our model predictions.
We conclude the fiducial model M4.0c is the best case, in which the predictions for both $\alpha_{153}^{608}(R)$ and $\nu J_{\nu}$
are in reasonable agreement with the observations.

Note that in previous studies including Paper I the integrated spectrum was often presented in the form of $J_{\nu}$ 
typically over four orders of magnitudes, 
so it gave much better visual impressions for the comparison between the predicted and the observed spectra.

\section{Summary}

Many of observed features of giant radio relics are thought to be explained by the shock acceleration model:
elongated shapes on scales of Mpc, radio spectral index
steepening toward the cluster center, and high polarization levels \citep{vanweeren10,stroe16}.
Among some remaining puzzles concerning the shock acceleration model,
in the case of the Sausage relic, we notice (1) the steep spectral curvature above GHz in the volume-integrated spectrum
\citep{stroe16}
and (2) the discrepancy between the X-ray based shock Mach number, 
$M_{\rm X}\approx 2.7$ and the radio based value, $M_{\rm radio}\approx 4.6$ \citep{akamatsu15,vanweeren10}.
To understand these features, in earlier studies we explored the reacceleration scenario, in which 
a weak shock with $M_s\approx 3$ propagates through a finite-size cloud of the ICM gas, containing a flat spectrum of
preexisting CRe \citep{kangryu16,kang16b}.
Considering the short cooling time of GeV electrons, however, it remains challenging to explain how to maintain 
such a flat population of high-energy electrons over a large preshock volume \citep{kang17}.

In this study, we explore an alternative model in which a shock of $M_s\approx 3-4$ sweeps through a preshock 
cloud containing low-energy 
fossil electrons and the electron aging is delayed by Fermi II acceleration by postshock turbulence.
Here preexisting CRe with $\gamma_e\lesssim 300$ provide only seed electrons to Fermi I process, so
the slope of the electron spectrum at the shock is determined by the sonic Mach number, 
i.e., $q=4M_{\rm s}^2/( M_{\rm s}^2-1)$.
This eliminates the unrealistic requirements for a flat power-law spectrum ($s=4.1$) with a high energy cutoff 
($\gamma_{\rm e,c}\approx 3-5\times 10^4$) adopted in \citet{kang16b} (Paper I).
Stochastic acceleration via transit time damping resonance off compressive MHD turbulence 
in the postshock region is adopted,
since the observed width of the Sausage relic is somewhat too broad to be explained sorely by the electron cooling
length (see the M4.0cN model in Figure 4).
We find that turbulent acceleration with $\tau_{\rm acc}\approx 10^8$~yr is required
in order to match the observed broad profiles of the radio flux density, $S_{\nu}(R)$,
of the Sausage relic.
We note such a strength of turbulence acceleration is similar to what is required to reproduce the radio flux
profiles of the Toothbrush radio relic \citep{kang17}.

Here we attempt to reproduce the observed profiles of $S_{\rm 153MHz}$, $S_{\rm 608MHz}$,
and the spectral index $\alpha_{153}^{608}$ as well as the volume-integrated
spectrum $J_{\nu}$ of the Sausage radio relic \citep{stroe16}.
In the best fitting fiducial model, M4.0c (see Table 1 for the model parameters), 
the spherical shock with the initial Mach number $M_{s,i}=4.0$ and the radius $r_{\rm s,i}=0.8\Mpc$ 
encounters the cloud of preexisting CRe and then sweeps out of the cloud after $t_{\rm exit}\approx 200$~Myr.
It turns out that the degree of the spectral steepening above GHz in $J_{\nu}$ strongly constrains
the duration, $t_{\rm obs}-t_{\rm exit}\approx 10$~Myr,
during which the shock propagates in the preshock region without fossil CRe.
At the time of observation, the model shock weakens to $M_{\rm s,obs}\approx3.2$ and the postshock temperature becomes
$kT_{\rm 2,obs}\approx 8.6$~keV, which are in reasonable agreements with X-ray observations \citep{akamatsu15}. 
Note that the M4.0c model does not reproduce the flattest observed index, $\alpha_{153}^{608}\approx 0.54$, at the relic edge,
which requires a much stronger shock with $M_{\rm s}\approx 6.9$.

As shown in Figures 3 and 4, the spectral index profile, $\alpha_{153}^{608}(R)$, provides the most
stringent constraints to the model parameters such as $M_{\rm s,i}$, $L_{\rm cloud}$, $B_1$, and $\tau_{\rm acc}$.
The amount of fossil low-energy CRe that can produce the observed radio flux density corresponds to the pressure ratio $N=P_{\rm CRe}/P_{\rm g}\approx 10^{-4}$,
which is dynamically insignificant.
Considering that the observational error in $J_{\nu}$ is about 10 \%, we could argue that the model predictions in Figure 5
($\log \nu J_{\nu}$ versus $\log \nu$) only marginally fit the observed integrated spectrum.

This study demonstrates that it is possible to explain most of the observed properties of the Sausage relic
by the shock reacceleration model with fossil relativistic electrons and an additional postshock Fermi II acceleration.
This scenario is consistent with the observational fact that
only a small fraction ($\sim 10\%$) of merging clusters host radio relics \citep{feretti12}.
Thus we favor the DSA reacceleration model in which radio relics are generated preferentially 
when merger-driven shocks encounter the regions containing preexisting low-energy CRe.

\acknowledgments{
This research was supported by Basic Science Research Program through the National Research Foundation of Korea (NRF) funded by the Ministry of Education (2014R1A1A2057940) and NRF grant (2016R1A5A1013277).
}


\end{document}